\documentclass[12pt,a4paper]{article}
\usepackage{amsfonts,latexsym}
\usepackage{amsmath,amssymb}
\usepackage{graphicx,color}

\numberwithin{equation}{section}

\renewcommand{\thefootnote}{\fnsymbol{footnote}}

\newcommand{\nn}{\nonumber}

\begin{document}
\vspace{12mm}

\begin{center}
{{{\Large {\bf Stability of Schwarzschild black holes \\ in fourth-order gravity revisited }}}}\\[10mm]

{Yun Soo Myung\footnote{e-mail address: ysmyung@inje.ac.kr}}\\[8mm]

{Institute of Basic Sciences and Department  of Computer Simulation, Inje University Gimhae 621-749, Korea\\[0pt]}

\end{center}
\vspace{2mm}

\begin{abstract}
We  revisit  the classical stability of Schwarzschild black hole
in fourth-order theories of gravity. On the contrary of the
stability of this black hole,  it turns out that the linearized
perturbations exhibit unstable modes featuring the
Gregory-Laflamme instability of five-dimensional black string.
This shows clearly an instability of black hole in fourth-order
gravity with $\alpha=-\beta/3$.

\end{abstract}
\vspace{5mm}

{\footnotesize ~~~~PACS numbers: 04.70.Bw, 04.50.Kd }


\vspace{1.5cm}

\hspace{11.5cm}{Typeset Using \LaTeX}
\newpage
\renewcommand{\thefootnote}{\arabic{footnote}}
\setcounter{footnote}{0}


\section{Introduction}

All black hole solutions found in Einstein gravity must pass the
stability test. A black hole solution should be stable against the
external perturbations because it stands as a physically realistic
object~\cite{Vish}. One decouples the linearized equations and then,
manages  to arrive at the  Schr\"odinger-type equations for physical
fields with their potentials. For non-rotating black holes, the
stable or unstable nature of black hole perturbation is determined
by the shape of the potential. If all potentials are positive for
whole range outside the event horizon, the black hole under the
consideration is stable.

However, for black holes found in massive gravity theories, the
method is not so straightforward to arrive at the stability of
black holes because they have more physical degrees of freedom
than two of a massless graviton.  Whitt~\cite{Whitt:1985ki} has
found  that provided both massive spin-0 and spin-2 are
non-tachyonic, the Schwarzschild black hole is classically stable
in fourth-order gravity. This implies that  initial perturbations
which are regular at infinity and on the future event horizon
cannot grow unboundedly with time.

Very recently, Babichev and Fabbri~\cite{Babichev:2013una} have
shown that massive linearized equation around the Schwarzschild
black hole in massive gravity and bimetric theories gives rise to
an instability of $l=0$ mode  by comparing it with the linearized
equation around  the black string in five dimensions. It turned
out that the bimetric black hole solution is unstable provided
$m'=m\sqrt{1+1/\kappa}$ satisfies a bound of $0<m'<\frac{{\cal
O}(1)}{2M_{\rm S}}$. In addition, the authors
~\cite{Brito:2013wya} have confirmed this result by considering a
more generic framework and extending $l=0$ mode to generic modes.

We wish to point out that the above two have the same linearized
equation when replacing the Ricci tensor perturbation $\delta
\tilde{R}_{\mu\nu}$ by the metric perturbation $h^{(-)}_{\mu\nu}$.
However, the above two give controversy  to each other. This is why
we will  revisit the issue of stability of Schwarzschild black hole
in fourth-order theories of gravity.  We find that
 the black hole in fourth-order gravity with $\alpha=-\beta/3$ is unstable
 since  $m_2=1/\sqrt{\beta}$ may satisfy a  bound of
$0<m_2<\frac{{\cal O}(1)}{2M_{\rm S}}$.

\section{Perturbation of  black holes in fourth-order gravity}
We start with the fourth-order gravity  theory given
by~\cite{Whitt:1985ki}
\begin{eqnarray}
S=\frac{1}{16 \pi}\int d^4 x\sqrt{-g} \Big[R-\alpha R^2-\beta
R_{\mu\nu}R^{\mu\nu}\Big] \label{Action}
\end{eqnarray} with two parameters $\alpha$ and $\beta$.
This  theory is renormalizable in Minkowski
spacetimes~\cite{stelle} and it describes 8 degrees of freedom
(DOF) [a massless spin-2 graviton with 2 DOF, a massive spin-2
graviton with 5 DOF, and a massive spin-0 with 1
DOF]~\cite{Myung:2011nn}. However, the massive graviton suffers
from having ghosts. We note that $f(R)$-gravity with $\beta=0$ has
3 DOF (a massless spin-2 graviton and a massive spin-0) without
ghost~\cite{Myung:2010rj}.

 The Einstein equation  takes the form
\begin{eqnarray} \label{equa1}
R_{\mu\nu}-2\alpha RR_{\mu\nu}&-&2\beta
R^{\rho\sigma}R_{\rho\mu\sigma\nu}-\beta \nabla^2
R_{\mu\nu}+(2\alpha+\beta)\nabla_\nu\nabla_\mu R \nonumber \\
&=&\frac{g_{\mu\nu}}{2}\Big[R-\alpha R^2-\beta
R_{\rho\sigma}R^{\rho\sigma}+(4\alpha+\beta)\nabla^2 R\Big].
\end{eqnarray}
It is well-known that Eq.(\ref{equa1}) provides  the Schwarzschild
black hole solution \begin{equation} \label{sch} ds^2_{\rm
S}=\bar{g}_{\mu\nu}dx^\mu
dx^\nu=-e^{\nu(r)}dt^2+e^{-\nu(r)}dr^2+r^2d\Omega^2_2
\end{equation}
with the metric function \begin{equation} \label{num}
e^{\nu(r)}=1-\frac{2M_{\rm S}}{r}.
\end{equation} In this case, the trace of (\ref{equa1}) becomes an
equation for the Ricci scalar
\begin{eqnarray}
2(3\alpha+\beta)\nabla^2 R+R=0.\label{eqR}
\end{eqnarray}
In the case of $\alpha=-\beta/3$, there is  no scalar
graviton ($R=0$)~\cite{LP}.

 In order to perform the stability analysis, we
usually introduce the metric perturbation around the black hole
\begin{eqnarray} \label{m-p}
g_{\mu\nu}=\bar{g}_{\mu\nu}+h_{\mu\nu}.
\end{eqnarray}
Hereafter we denote the background quantities with the
``overbar''. However, in fourth-order gravity, the metric
perturbation (\ref{m-p}) leads to the fourth-order differential
equation~\cite{GT} which is almost  impossible to be solved.
Instead, we consider the resulting perturbation in Ricci tensor,
$\delta R_{\mu\nu}(h)$ whose representation is given by
\begin{eqnarray}
\delta
R_{\mu\nu}(h)&=&\frac{1}{2}\Big(\bar{\nabla}^{\rho}\bar{\nabla}_{\mu}h_{\nu\rho}+
\bar{\nabla}^{\rho}\bar{\nabla}_{\nu}h_{\mu\rho}-\bar{\nabla}^2h_{\mu\nu}-\bar{\nabla}_{\mu}
\bar{\nabla}_{\nu}h\Big).\label{lRmun}
\end{eqnarray}
The linearized equation for Ricci tensor takes the second-order
differential form
\begin{eqnarray}
\Bigg[\Big\{\beta \bar{g}_{\mu\rho}\bar{g}_{\nu\sigma}&+&\Big(2\alpha+\frac{\beta}{2}\Big)\bar{g}_{\rho\sigma}\bar{g}_{\mu\nu}\Big\}
\bar{\nabla}^2-\bar{g}_{\mu\nu}(2\alpha+\beta)\bar{\nabla}_\rho\bar{\nabla}_\sigma, \nonumber \label{lRmunu}\\
&+&\Big(2\beta
\bar{R}_{\mu\rho\nu\sigma}-\bar{g}_{\mu\rho}\bar{g}_{\nu\sigma}+\frac{1}{2}
\bar{g}_{\rho\sigma}\bar{g}_{\mu\nu}\Big)\Bigg]\delta
R^{\mu\nu}=0.\label{lR}
\end{eqnarray}
Taking the trace of (\ref{lRmunu})
 leads to the linearized Ricci
scalar equation
\begin{equation} \label{rseq}
\Big(\bar{\nabla}^2 -m^2_0\Big)\delta R=0,\end{equation} where the
spin-0 mass $m_0^2$ is given by
\begin{equation}
m^2_0=-\frac{1}{2(3\alpha+\beta)}.
\end{equation}
Eq.(\ref{rseq}) is consistent with that obtained from  varying
(\ref{eqR}) directly. Here we may choose  $2(3\alpha+\beta)<0$ and
$\beta>0$ for $m^2_0>0$ and $m_2^2=1/\beta>0$. In the case of
$\alpha=-\beta/3$, this scalar graviton is decoupled from the
theory, implying that $\delta R=0$.

To simplify the stability analysis, we decompose $\delta R_{\mu\nu}$
into the trace and trace-free  parts as
\begin{equation} \label{liri}
\delta R_{\mu\nu}=\frac{\delta R}{4} \bar{g}_{\mu\nu}+\delta
\tilde{R}_{\mu\nu},\end{equation} where
\begin{equation} \label{tln}\bar{g}^{\mu\nu}\delta
\tilde{R}_{\mu\nu}=0.\end{equation} Then, the linearized equation
of (\ref{lR}) becomes
\begin{eqnarray}
&& \Bigg[\bar{g}_{\mu\rho}\bar{g}_{\nu\sigma}\Big(\bar{\nabla}^2-\frac{1}{\beta}\Big) +2\bar{R}_{\mu\rho\nu\sigma}\Bigg]\delta \tilde{R}^{\mu\nu}\nn\\
&&
+\Bigg[\Big(\frac{2\alpha}{\beta}+\frac{3}{4}\Big)\bar{g}_{\rho\sigma}\bar{\nabla}^2-
\Big(\frac{2\alpha}{\beta}+1\Big)\bar{\nabla}_{\rho}\bar{\nabla}_{\sigma}+\frac{1}{4\beta}\bar{g}_{\rho\sigma}\Bigg]\delta
R=0.\label{leq1}
\end{eqnarray}
Imposing  $\beta=0$ [after multiplying  (\ref{leq1}) by $\beta$], it
reduces to the linearized equation around the Schwarzschild black
hole found in  $f(R)$-gravity~\cite{Myung:2011ih}.

At this stage, we follow the method used in~\cite{Whitt:1985ki} to
perform the  stability analysis by separating (\ref{leq1}) into two
parts: one is the equation for $\delta \tilde{R}^{\mu\nu}$ as
\begin{eqnarray} \label{masseq}
&& \Big[\bar{g}_{\mu\rho}\bar{g}_{\nu\sigma}\bar{\nabla}^2
+2\bar{R}_{\mu\rho\nu\sigma}-\frac{1}{\beta}\bar{g}_{\mu\rho}\bar{g}_{\nu\sigma}\Big]\delta
\tilde{R}^{\mu\nu}=0.\label{leq2}
\end{eqnarray}
The other may be  given by taking the trace of (\ref{leq1}), which
leads to (\ref{rseq}). Hence, we do not consider (\ref{rseq})
further because it is stable against the perturbation provided
$m^2_0>0$. Actually, this separation  was proposed to  decouple the
massive spin-2 equation from the massive spin-0 equation even though
its validity is not yet proved.  We may set $\delta R$ to be zero in
favor of obtaining equation (\ref{leq2}) for $\delta
\tilde{R}^{\mu\nu}$ which is the key ingredient for describing the
massive spin-2 graviton. However, this is not the case unless
$\alpha=-\beta/3$. Hence, for $\alpha=-\beta/3(\delta
R=0)$~\cite{LP}, Eq.(\ref{lR}) could reduce to  Eq.(\ref{leq2})
which becomes a correct linearized equation for $\delta
\bar{R}_{\mu\nu}$. We note that the condition of $\alpha=-\beta/3$
kills a massive spin-0 graviton with mass $m^2_0$.

Our action (\ref{Action}) reveals ghosts when performing the metric
perturbation  $h_{\mu\nu}$ around the Minkowski spacetimes
($M^4$)~\cite{stelle}. Explicitly, for choosing $\alpha=-\beta/3$,
its Lagrangian takes the form of ${\cal L} = \sqrt{-g}[R-\beta
(R_{\mu\nu}^2-R^2/3)]$. Although the $\beta$-term of providing
massive graviton  with mass $m_2^2=1/\beta$ improves the ultraviolet
divergence, it induces ghost
excitation~\cite{Barth:1983hb,Myung:2011nn}
\begin{equation}
\frac{1}{p^2}-\frac{1}{p^2+m^2_2}
\end{equation}
which spoils the unitarity but provides a healthy massless graviton.
The price one has to pay for making the theory renormalizable is the
loss of unitarity. If one is interested in  the massive graviton
propagation only, it would be better  go to the three dimensional
new massive gravity~\cite{Bergshoeff:2009hq} where the massive
graviton propagates without ghost because of the sign change in
$-R$~\cite{Nakasone:2009bn}. However, this is possible in three
dimensions only, but  it provides a massless graviton with negative
norm state in four dimensions
\begin{equation}
-\frac{1}{p^2}+\frac{1}{p^2+m^2_2}.
\end{equation}   Even though
the ghosts (massless graviton) are unavoidable, we may choose `$-R$'
instead of $R$ to have a  healthy massive graviton in the
metric-perturbation theory of the fourth-order gravity.  We insist
that from (\ref{masseq}), the Ricci-tensor perturbation $\delta
\tilde{R}^{\mu\nu}$ has  the massive graviton propagator  shown as
\begin{equation} \frac{1}{p^2+m^2_2}
\end{equation} around the Minkowski spacetimes.

It is proposed  that the ghost issue of the  fourth-order gravity
persists in the Schwarzschild background.  As was pointed out for
the Kaluza-Klein compactifications on $M^4\times
S^2$~\cite{RandjbarDaemi:1982hi} and $M^4\times S^2\times S^2\times
S^2$~\cite{Myung:1986da}, the stability condition is achieved
provided  that the on-shell amplitude is free from tachyons and
ghosts upon plugging the external sources.

In the case of black hole stability, a conventional method of
determining the stability is to solve the second-order linearized
equation by choosing even-and odd-parity perturbations under the
Regge-Wheeler gauge for a massless graviton, which leads to two
Schr\"odinger equations: Regge-Wheeler equation~\cite{Regge:1957td}
and Zerilli equation~\cite{Zeri}. One may conclude that the
Schwarzschild black hole is stable because their potentials are
positive definite for the whole region outside the black hole,
implying that there is no exponentially growing modes with respect
to time~\cite{Vish}. If one wants to perform the stability analysis
of the Schwarzschild black hole in the fourth-order gravity, one
should  encounter  difficulty in handling fourth-order derivatives
in the linearized equation which includes ghost propagations.  Up to
now, we do not know how to perform the stability  analysis of black
hole in the metric-perturbation theory of the fourth-order gravity.
This is why we introduce the Ricci-tensor perturbation theory which
hides the appearance of  fourth-order derivatives.

\section{Re-analysis of the  black hole
stability in fourth-order gravity}

In  Einstein gravity, its linearized equation takes a simple form as
\begin{equation} \label{leeq}
\delta R_{\mu\nu}(h)=0.
\end{equation} Then, the metric perturbation $h_{\mu\nu}$ is
classified depending on the transformation properties under parity,
namely odd  and even. Using the Regge-Wheeler~\cite{Regge:1957td}
and Zerilli gauge~\cite{Zeri}, one obtains two distinct
perturbations: odd and even perturbations. For odd parity,
$h^o_{\mu\nu}$ includes  two off-diagonal components $h_0$ and $h_1$
as
\begin{eqnarray}
h^o_{\mu\nu}=\left(
\begin{array}{cccc}
0 & 0 & 0 & h_0(r) \cr 0 & 0 & 0 & h_1(r) \cr 0 & 0 & 0 & 0 \cr
h_0(r) & h_1(r) & 0 & 0
\end{array}
\right) e^{-ik t}\sin\theta\frac{dp_{l}}{d\theta} \,, \label{oddp}
\end{eqnarray}
while for even parity, the metric tensor takes the form with four components $H_0,~H_1,~H_2,$ and $K$ as
\begin{eqnarray}
h^e_{\mu\nu}=\left(
\begin{array}{cccc}
H_0(r) e^{\nu(r)} & H_1(r) & 0 & 0 \cr H_1(r) & H_2(r) e^{-\nu(r)} &
0 & 0 \cr 0 & 0 & r^2 K(r) & 0 \cr 0 & 0 & 0 & r^2\sin^2\theta K(r)
\end{array}
\right) e^{-ikt}p_{l} \,, \label{evenp}
\end{eqnarray}
where $p_l$ is Legendre polynomial with angular momentum number
$l$ and $e^{\nu(r)}$ is given by (\ref{num}). For $l=0$
($s$-mode), $h^e_{\mu\nu}$ survives only because of $p_0=1$. This
mode is responsible for indicating  the instability of the black
hole in fourth-order gravity.  Plugging $h^o_{\mu\nu}$ and
$h^e_{\mu\nu}$ into the linearized Einstein equation (\ref{leeq})
leads to Regge-Wheeler and Zerilli equations which describe two
physical DOF propagating on the black hole spacetimes. Their
non-negative potentials guarantee the stability of the
Schwarzschild black hole in the Einstein gravity~\cite{chan}.

In fourth-order gravity with $\alpha=-\beta/3$, however,  we have to solve (\ref{masseq})
which is rewritten as
\begin{equation} \label{massteq}
\bar{\nabla}^2 \delta
\tilde{R}_{\mu\nu}+2\bar{R}_{\rho\mu\sigma\nu}\delta
\tilde{R}^{\rho\sigma}=m^2_2 \delta \tilde{R}_{\mu\nu}.
\end{equation}
At this stage, we stress again that  (\ref{massteq}) is considered
as the second-order equation with respect to $\delta
\tilde{R}_{\mu\nu}$, but not the fourth-order equation for
$h^{o/e}_{\mu\nu}$. In  massive gravity theory, there is
correspondence between linearized Ricci tensor $\delta R_{\mu\nu}$
and Ricci spinor $\Phi_{ABCD}$ when using the Newman-Penrose
formalism~\cite{Newman:1961qr}.  Here the null real tetrad is
necessary to specify polarization modes of massive graviton, as the
four-dimensional massive gravity requires  null complex tetrad to
specify six  polarization modes~\cite{Eardley:1974nw}.  This implies
that in fourth-order gravity theory, one may take linearized Ricci
tensor $\delta R_{\mu\nu}$ (\ref{liri}) with 6 DOF
[$\delta\tilde{R}_{\mu\nu}$ with 5 DOF] as physical observables,
instead of linearized metric tensor $h_{\mu\nu}$ in Einstein
gravity~\cite{Moon:2011gg}. Also, we have the tracelessness
(\ref{tln}) of $\delta \tilde{R}^{\mu}~_\mu=0$ and the
transversality of $\bar{\nabla}^\mu \delta \tilde{R}_{\mu\nu}=0$
from the contracted Bianchi identities.

 According to
Ref.~\cite{Whitt:1985ki} based on (\ref{leq2}), the author has
concluded that the Schwarzschild black hole is classically stable in
fourth-order gravity, even though he mentioned a possibility of a
static $s$-mode ($\Omega=ik=0,l=0$) instability for $m_2^2=1/\beta
<0.19M^2_{\rm S}=1/(2.294M_{\rm S})^2$.

 Now we are in a position to revisit this conclusion of stability on
 the Schwarzschild black hole in fourth-order gravity with $\alpha=-\beta/3$.
For this purpose, we stress to note that (\ref{massteq}) takes the
following  form when replacing $\delta \tilde{R}_{\mu\nu}$ and
$m^2_2$ by $h^{(-)}_{\mu\nu}$ and $m'^2$:
\begin{equation} \label{mgbt}
\bar{\nabla}^2
h^{(-)}_{\mu\nu}+2\bar{R}_{\rho\mu\sigma\nu}h^{(-)\rho\sigma}=m'^2
h^{(-)}_{\mu\nu}
\end{equation}
together with the transverse-traceless condition of $\nabla^\mu
h^{(-)}_{\mu\nu}=0$ and $h^{(-)\mu}~_\mu=0$. Surely, Eq.(\ref{mgbt})
is the linearized  equation around the Schwarzschild black hole in
massive gravity and bimetric theories~\cite{Babichev:2013una}. The
Gregory-Laflamme ($l=0$ mode) instability was recently used to
point out the instability of the Schwarzschild black hole in massive
gravity and bimetric theories. Considering the
perturbations~\cite{Gregory:1993vy}
\begin{equation}
\left(
  \begin{array}{cc}
   h^{(4)}_{\mu\nu} & h_{\mu z} \\
  h_{z\nu} & h_{zz} \\
  \end{array}
\right)
\end{equation}
 around the
five-dimensional black string
\begin{equation}
ds^2_{\rm 5bs}=ds^2_{\rm S}+dz^2,
\end{equation}
$h^{(4)}_{\mu\nu}$ satisfies the massive spin-2 equation
(\ref{mgbt}) with $m'^2=\mu^2$ through  $e^{i\mu z}$. It is pointed
out that unstable modes of $h^e_{\mu\nu}$ (\ref{evenp})$\to
h^{(-)}_{\mu\nu}$ (\ref{mgbt})  with $l=0$ and $k=i\Omega$ which are
regular at the future event horizon were found within range of
$0<m'<\frac{{\cal O}(1)}{2M_{\rm S}}$~\cite{Babichev:2013una}.
Furthermore, the authors~\cite{Brito:2013wya} have very recently
shown that the Schwarzschild black holes are generally unstable
against spherically symmetric perturbations in massive gravity and
bimetric theories.

Similarly, we find that unstable modes of $h^e_{\mu\nu}$
(\ref{evenp})$\to \delta \tilde{R}^e_{\mu\nu}$ (\ref{massteq})
with $l=0$ and $k=i\Omega$
 where are regular at the future event horizon  were found within range of $0<m_2<\frac{{\cal O}(1)}{2M_{\rm
S}}$.

Consequently, the Schwarzschild black hole is unstable against the
perturbations $\delta \tilde{R}_{\mu\nu}$ in fourth-order gravity
with $\alpha=-\beta/3$ because (\ref{massteq}) becomes (\ref{mgbt})
when replacing $\delta \tilde{R}_{\mu\nu}$ and $m^2_2$ by
$h^{(-)}_{\mu\nu}$ and $m'^2$.  We note that the instability of
Schwarzschild black hole is not proved in fourth-order gravity with
arbitrary $\alpha$ and $\beta$ because its linearized equation is
given by the complicated form (\ref{leq1}) which is surely a coupled
second-order equation for $\delta \tilde{R}_{\mu\nu}$ and $\delta
R$.

Finally, we would like to mention  that Whitt~\cite{Whitt:1985ki}
has proposed initially  a static ($k=0$) $s$-mode ($l=0$)
instability for $m_2^2=1/\beta <0.19M^2_{\rm S}=1/(2.294M_{\rm
S})^2$ because it indicates a bifurcation of $l=0$ solution of the
fourth-order gravity theory. This implies that there is another
family of $l=0$ solution, which is identical with the Schwarzschild
solution (\ref{sch}) at a critical mass $0.44\beta$. His bound of
$0<m_2 <\frac{1}{2.294M_{\rm S}}$ is similar to  the instability
bound of $0<m'<\frac{{\cal O}(1)}{2M_{\rm S}}$. However, he has
shown that
 a static
($k=0$) $s$-mode ($l=0$) does not become an unstable mode for $0<m_2
<\frac{1}{2.294M_{\rm S}}$ (see Appendix E in~\cite{Whitt:1985ki}).
Actually, the static perturbation has nearly nothing to do with the
black hole stability.

Importantly, he has insisted that for non-static $(k\not=0)$
$s$-mode, there is no unstable perturbations. His flaw seems to
arise from the fact that as was shown in Appendix D, he treated the
$l=0$ case on equal footing with the  $l=1$ case. In the case of
massive spin-2 theory, the first instability arises from the $l=0$
perturbation. In other words, he did not solve the $l=0$
perturbation equation correctly. Before a seminal work of the black
string stability~\cite{Gregory:1993vy}, one does not  realize that
the $l=0$ perturbation equation reveals    an unstable mode.

 Furthermore, he
has used the simple form (\ref{massteq}) for arbitrary $\alpha$ and
$\beta$ without proving its validity. In this case, (\ref{leq1})
cannot reduce to (\ref{massteq}). Thus, his choice of
(\ref{massteq}) is not legitimate. The validity of (\ref{massteq})
is not proved unless $\alpha=-3\beta$. Our conclusion on the
instability of black hole is suitable for $\alpha=-3\beta$ case, but
not for $\alpha \not=-3\beta$, where a scalar graviton $\delta R$
with mass $m_0^2$ propagates around the Schwarzschild black hole
spacetimes. Frankly,  we have no idea on solving the complicated
equation (\ref{leq1}).

 \vspace{1cm}

{\bf Acknowledgments}

This work was supported by the 2013 Inje University Research Grant.

\end{document}